\documentclass[11pt]{article}
\usepackage[T1]{fontenc}
\usepackage[utf8]{inputenc}
\usepackage{times,cite,amsmath,amssymb,amsfonts,marvosym,wasysym,graphicx,rotating}
\usepackage[font=sf,labelfont=bf]{caption}
\DeclareCaptionLabelSeparator{bar}{ | }
\usepackage[letterpaper,top=1.5cm,bottom=2cm,left=2cm,right=2cm]{geometry}
\usepackage[colorlinks=true]{hyperref}

\newcommand{\spacing}[1]{\renewcommand{\baselinestretch}{#1}\large\normalsize}
\spacing{1.5}

\begin{document}

\newenvironment{affiliations}{%
\setcounter{enumi}{1}%
\setlength{\parindent}{0in}%
\slshape\sloppy%
\begin{list}{\upshape$^{\arabic{enumi}}$}{%
\usecounter{enumi}%
\setlength{\leftmargin}{0in}%
\setlength{\topsep}{0in}%
\setlength{\labelsep}{0in}%
\setlength{\labelwidth}{0in}%
\setlength{\listparindent}{0in}%
\setlength{\itemsep}{0ex}%
\setlength{\parsep}{0in}%
}
}{\end{list}\par\vspace{12pt}}

\renewenvironment{abstract}{%
\setlength{\parindent}{0in}%
\setlength{\parskip}{0in}%
\bfseries%
}{\par\vspace{-6pt}}

\newenvironment{methods}{%
\section*{Methods}%
\setlength{\parskip}{12pt}%
}{}

\newenvironment{addendum}{%
\setlength{\parindent}{0in}%
\small%
\begin{list}{Acknowledgements}{%
\setlength{\leftmargin}{0in}%
\setlength{\listparindent}{0in}%
\setlength{\labelsep}{0em}%
\setlength{\labelwidth}{0in}%
\setlength{\itemsep}{12pt}%
\let\makelabel\addendumlabel}
}
{\end{list}\normalsize}
\newcommand*{\addendumlabel}[1]{\textbf{#1}\hspace{1em}}

\begin{flushleft}
{\Large\sffamily\bfseries Planar parallel phonon Hall effect and local symmetry breaking}
\end{flushleft}

\begin{flushleft}
Quentin Barthélemy$^{1}$\Letter, Étienne Lefrançois$^{1}$, Lu Chen$^{1}$, Ashvini Vallipuram$^{1}$, Katharina M. Zoch$^{2}$, Cornelius Krellner$^{2}$, Pascal Puphal$^{3}$\Letter~\& Louis Taillefer$^{1,4}$\Letter
\end{flushleft}

\begin{affiliations}
\item Institut Quantique, Département de physique \& RQMP, Université de Sherbrooke, Sherbrooke, Québec, Canada
\item Physikalisches Institut, Goethe Universität Frankfurt, Frankfurt am Main, Germany
\item Max Planck Institute for Solid State Research, Stuttgart, Germany
\item Canadian Institute for Advanced Research, Toronto, Ontario, Canada
\item[\Letter] e-mail: quentin.barthelemy@usherbrooke.ca; p.puphal@fkf.mpg.de; louis.taillefer@usherbrooke.ca
\end{affiliations}

\begin{abstract}
Y-kapellasite [Y$\mathbf{_{3}}$Cu$\mathbf{_{9}}$(OH)$\mathbf{_{19}}$Cl$\mathbf{_{8}}$] is a frustrated antiferromagnetic insulator which remains paramagnetic down to a remarkably low Néel temperature of about $\mathbf{2}$~K. Having studied this material in the paramagnetic regime, in which phonons are the only possible heat carriers, we report the observation of a planar parallel thermal Hall effect coming unambiguously from phonons. This is an advantage over the Kitaev quantum spin liquid candidates {\boldmath$\alpha$}-RuCl$\mathbf{_{3}}$ and Na$\mathbf{_{2}}$Co$\mathbf{_{2}}$TeO$\mathbf{_{6}}$ where in principle other heat carriers can be involved\cite{Yokoi_Science_2021,Czajka_NM_2022,Takeda_PRR_2022,Chen_arXiv_2023_a}. As it happens, Y-kapellasite undergoes a structural transition attributed to the positional freezing of a hydrogen atom below about $\mathbf{33}$~K. Above this transition, the global crystal symmetry forbids the existence of a planar parallel signal -- the same situation as in Na$\mathbf{_{2}}$Co$\mathbf{_{2}}$TeO$\mathbf{_{6}}$ and cuprates\cite{Takeda_PRR_2022,Chen_arXiv_2023_a,Chen_arXiv_2023_b}. This points to the notion of a local symmetry breaking at the root of the phonon Hall effect. In this context, the advantage of Y-kapellasite over Na$\mathbf{_{2}}$Co$\mathbf{_{2}}$TeO$\mathbf{_{6}}$ (with high levels of Na disorder and stacking faults) and cuprates (with high levels of disorder coming from dopants and oxygen vacancies) is its clean structure, where the only degree of freedom available for local symmetry breaking is this hydrogen atom randomly distributed over six equivalent positions above $\mathbf{33}$~K. This provides a specific and concrete case for the general idea of local symmetry breaking leading to the phonon Hall effect in a wide range of insulators.
\end{abstract}

\clearpage

\begin{figure}[t!]\centering
\includegraphics[width=0.45\textwidth]{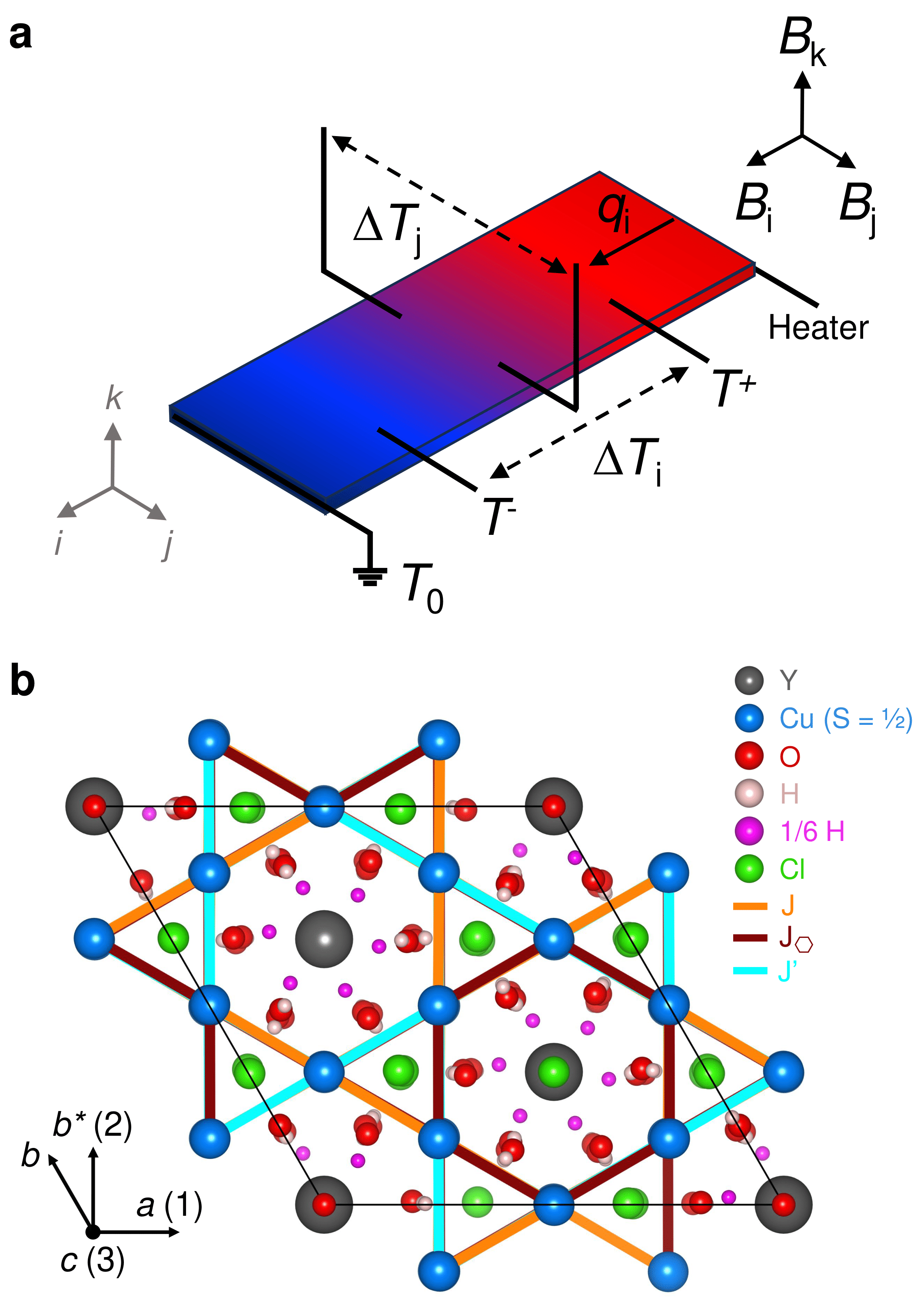}
\caption{\label{Fig1}\textbf{Thermal transport setup, structure and magnetic model of Y-kapellasite.} \textbf{a} Sketch of the thermal transport measurement setup showing the directions of the heat flux $q_{\mathrm{i}}$, the longitudinal thermal gradient $\Delta T_{\mathrm{i}}$, the thermal Hall gradient $\Delta T_{\mathrm{j}}$ and the applied magnetic field $B_{\mathrm{x}}$, with $\mathrm{x}\in\{i,j,k\}$. $(i,j,k)$ is an orthonormal basis. In a conventional thermal Hall effect configuration, the magnetic field is applied fully along $k$. In a planar parallel thermal Hall effect configuration, the magnetic field is applied fully along $i$. In a planar orthogonal thermal Hall effect configuration, the magnetic field is applied fully along $j$. \textbf{b} Crystal structure of Y-kapellasite [Y$_{3}$Cu$_{9}$(OH)$_{19}$Cl$_{8}$], with space group $R\overline{3}$ ($148$), deduced from neutron diffraction measurements at $40$~K (above $T_{\mathrm{S}1}$) and viewed along the $c$ axis\cite{Chatterjee_PRB_2023}. The orthonormal lattice vectors $a$, $b^{\star}$ and $c$ are numbered $1$, $2$ and $3$. Note the $1/6$ partial occupation of the inter-plane hydrogen positions (highlighted in pink). The three main antiferromagnetic couplings $J\simeq J_{\hexagon}\simeq140$~K and $J^{'}\simeq63$~K, indicated by thick coloured bonds, define an anisotropic nearest neighbour Heisenberg model on the slightly distorted copper kagome lattice.}
\end{figure}

Rather unexpectedly, C. Strohm and colleagues discovered the phonon Hall effect (PHE) through conventional thermal Hall effect measurements in a paramagnetic dielectric garnet, in which phonons are the only possible heat carriers\cite{Strohm_PRL_2005}. Given a longitudinal thermal gradient $\Delta T_{\mathrm{i}}$ produced by the heat flux $q_{\mathrm{i}}$ set at one end of the sample along the direction $i$, an orthogonal temperature gradient $\Delta T_{\mathrm{j}}$ develops along the direction $j$ when a magnetic field $B_{\mathrm{k}}$ is applied \emph{fully normal} to the $(ij)$ plane along the direction $k$, originating in a finite $Q_{\mathrm{ijk}}$ Righi-Leduc tensor component: $q_{\mathrm{i}}=Q_{\mathrm{ijk}}\Delta T_{\mathrm{j}}B_{\mathrm{k}}$, see Fig.~\ref{Fig1}\textbf{a}. In a conventional metal, where electrons carry heat in tandem with phonons, the thermal Hall conductivity $\kappa_{\mathrm{ij}}$ naturally includes a sizeable Lorentz force-like contribution, directly related to the electrical Hall effect through the Wiedemann-Franz law. Conversely, observing a finite $\kappa_{\mathrm{ij}}$ in an insulator, where all possible heat carriers are necessarily neutral, is quite counter-intuitive.

Over the past few years, a number of theoretical and experimental studies focusing on magnetic insulators highlighted that phonons are not the only possible neutral heat carriers and that collective spin excitations such as magnons -- conventional spin waves -- may also generate a thermal Hall effect\cite{Kitaev_AP_2006,Katsura_PRL_2010,Onose_Science_2010,Hirschberger_PRL_2015,Gao_SciPost_2020,Teng_PRR_2020,Chern_PRL_2021}. Naturally, the coupling of these magnetic quasiparticles to the magnetic field appears less cryptic. A tantalising aspect is that most studies conducted on frustrated quantum magnets reported a sizeable thermal Hall signal presented as compelling evidence for the emergence of long-sought exotic spin excitations while assuming a marginal or null PHE. 

The most recent and salient examples are those of two Kitaev quantum spin liquid candidates, namely $\alpha$-RuCl$_{3}$ and Na$_{2}$Co$_{2}$TeO$_{6}$, in which a startling planar parallel thermal Hall effect was detected when a magnetic field $B_{\mathrm{i}}$ is applied \emph{fully parallel} to the heat flux\cite{Yokoi_Science_2021,Czajka_NM_2022,Takeda_PRR_2022,Chen_arXiv_2023_a}, see Fig.~\ref{Fig1}\textbf{a}. Until now, this finite $Q_{\mathrm{iji}}$ was systematically attributed to unconventional magnetic edge states rather than phonons. In particular, regarding the highly scrutinised $\alpha$-RuCl$_{3}$, there is a heated debate to decide between the two scenarios proposed so far: chiral Majorana fermions from the gapped Kitaev quantum spin liquid\cite{Yokoi_Science_2021}, expected to yield a quantized temperature dependence of $\kappa_{\mathrm{ij}}$, versus topological magnons\cite{Czajka_NM_2022}, expected to yield a steeper temperature dependence, typical of bosons.

Yet, given that phonons contribute to the conventional thermal Hall effect in these two materials\cite{Lefrancois_PRX_2022,Gillig_arXiv_2023}, one wonders if they also contribute to the planar parallel thermal Hall effect. If so, this would cast doubt on the putative evidence for exotic spin excitations.

In our recent studies of Na$_{2}$Co$_{2}$TeO$_{6}$\cite{Chen_arXiv_2023_a} and of the Mott insulating (antiferromagnetic) cuprate Nd$_{2-\mathrm{x}}$Ce$_{\mathrm{x}}$CuO$_{4}$ (with $\mathrm{x}=0.04$)\cite{Chen_arXiv_2023_b}, we argued that the similar temperature dependence of the phonon-dominated longitudinal thermal conductivity and the planar parallel thermal Hall conductivity is a strong indication that phonons do contribute to the planar parallel thermal Hall effect. To support our interpretations, it is now crucial to provide an unambiguous observation of the planar parallel phonon Hall effect in a material where phonons are the only possible heat carriers.

Beyond the nature of the involved heat carriers, the symmetry requirements for the planar parallel thermal Hall effect constitute a key issue. In Na$_{2}$Co$_{2}$TeO$_{6}$ for instance, it is forbidden by the global crystal symmetries of the $P6_{3}22$ space group even though finite signals were reported by two independent groups, using samples from different sources\cite{Takeda_PRR_2022,Chen_arXiv_2023_a}. There is nonetheless a striking difference in magnitude and temperature dependence between the two sets of results, which suggests that mechanisms related to the sample quality and history, e.g., defects and domains, whether structural or magnetic, are responsible for the planar parallel thermal Hall effect. In Nd$_{2-\mathrm{x}}$Ce$_{\mathrm{x}}$CuO$_{4}$ (with $\mathrm{x}=0.04$), it is also forbidden by the global crystal symmetries of the $I4/mmm$ space group\cite{Chen_arXiv_2023_b}. We thus wonder if local symmetry breaking, around positional disorder or extrinsic defects, is at the root of the planar parallel thermal Hall effect.

In the present study, we investigate the thermal Hall effect in clean, phase pure single crystals of Y-kapellasite [Y$_{3}$Cu$_{9}$(OH)$_{19}$Cl$_{8}$], a kagome-based insulating frustrated antiferromagnet which does not display quantum spin liquid physics nor exotic spin excitations down to the lowest temperatures reached experimentally\cite{Chatterjee_PRB_2023}. While it undergoes two structural transitions at $T_{\mathrm{S}1}\simeq33$~K and $T_{\mathrm{S}2}\simeq13$~K, it remains a simple paramagnet down to a very low Néel temperature $T_{\mathrm{N}}\simeq2$~K above which phonons are the only possible heat carriers. In the high-temperature crystal structure (above $T_{\mathrm{S}1}$) defined by the space group $R\overline{3}$ ($148$), the inter-plane hydrogen is randomly distributed over six equivalent positions, see Fig.~\ref{Fig1}\textbf{b}. The transition at $T_{\mathrm{S}2}$ is still to be clarified but the transition at $T_{\mathrm{S}1}$ is attributed to the positional freezing of this atom and potentially leads to a crystal structure defined by the space group $P1$ with three different twin domains. This system thus provides a unique platform in which one can investigate the PHE and correlate symmetry considerations with the temperature dependence of specific thermal transport coefficients.

We performed thermal transport measurements on three high-quality single crystals of Y-kapellasite, labelled S$1$, S$2$ and S$3$. We focused on five distinct configurations listed in Table~\ref{Table1} so as to assess the $Q_{123}$ (conventional $\kappa_{12}$, in S$1$), $Q_{213}$ (conventional $\kappa_{21}$, in S$2$), $Q_{212}$ (planar parallel $\kappa_{21}$, in S$2$), $Q_{211}$ (planar orthogonal $\kappa_{21}$, in S$2$) and $Q_{321}$ (conventional $\kappa_{32}$, in S$3$) Righi-Leduc tensor components, where $1$, $2$ and $3$ respectively denote the $a$, $b^{\star}$ and $c$ orthonormal lattice vectors of the $R\overline{3}$ structure, see Fig.~\ref{Fig1}\textbf{b}.

\begin{figure}[t!]\centering
\includegraphics[width=\textwidth]{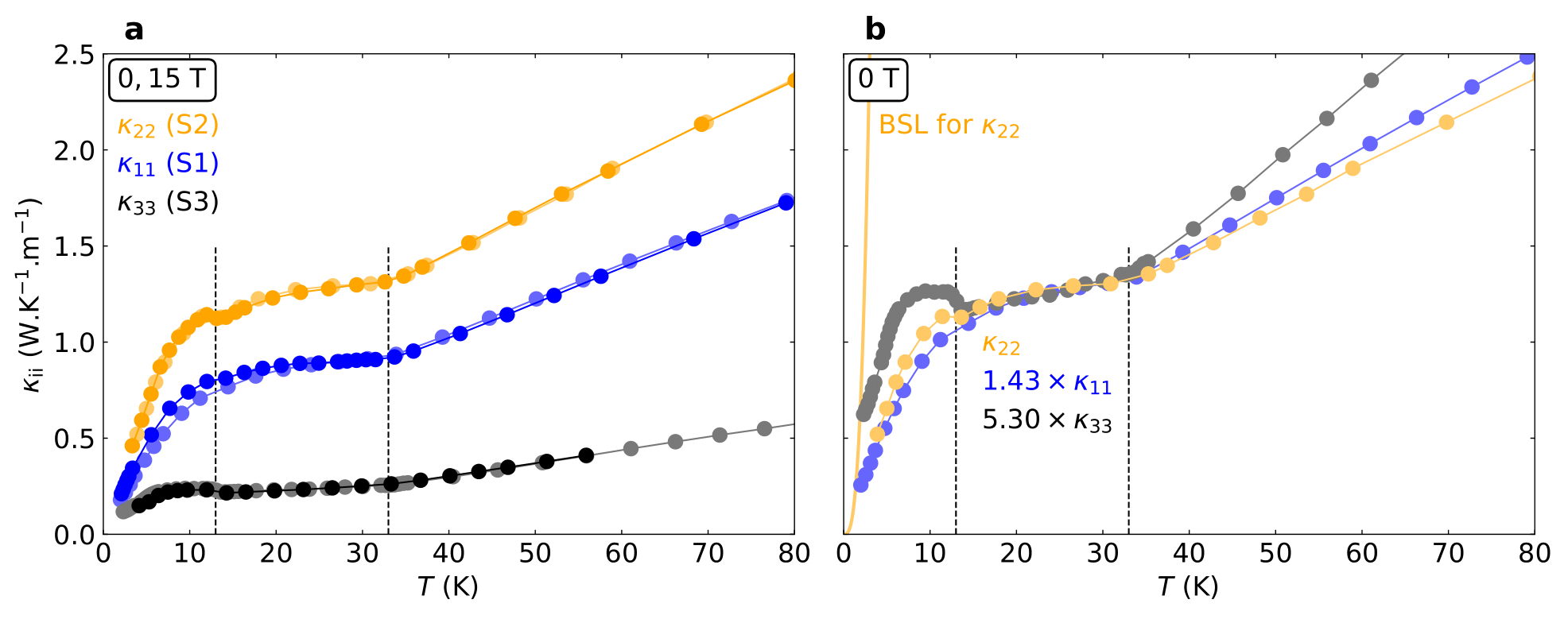}
\caption{\label{Fig2}\textbf{Phonon longitudinal thermal conductivity in Y-kapellasite.} \textbf{a} Temperature dependence of the three longitudinal thermal conductivities $\kappa_{11}$ (along $a$, measured on sample S$1$), $\kappa_{22}$ (along $b^{\star}$, measured on sample S$2$) and $\kappa_{33}$ (along $c$, measured on sample S$3$). Zero field data, in light colours, are plotted against $B_{3}=15$~T data, in dark colours (magnetic field applied along $c$). \textbf{b} The zero field data are reproduced from \textbf{a} with a multiplicative factor applied to $\kappa_{11}$ ($1.43$) and $\kappa_{33}$ ($5.30$) for a perfect match with $\kappa_{22}$ at $25$~K. The thick orange curve indicates the estimate of the low-temperature boundary scattering limit for $\kappa_{22}$. The two vertical black dashed lines in \textbf{a} and \textbf{b} indicate the structural transitions at $T_{\mathrm{S}1}\simeq33$~K and $T_{\mathrm{S}2}\simeq13$~K.}
\end{figure}

First examining the longitudinal thermal conductivities associated with the three corresponding heat flux directions, $\kappa_{11}$ (in S$1$), $\kappa_{22}$ (in S$2$) and $\kappa_{33}$ (in S$3$) as displayed in Fig.~\ref{Fig2}, we obtain confirmation that there are no other heat carriers than phonons at all considered temperatures from $80$ down to $2$~K. The temperature dependence of the three $\kappa_{\mathrm{ii}}$ is similar in shape, with clear anomalies at $T_{\mathrm{S}1}$ and $T_{\mathrm{S}2}$ reflecting the two structural transitions, see Fig.~\ref{Fig2}\textbf{a}. Actually, the three curves are almost identical up to a multiplicative factor which includes potential variations in crystalline quality and geometric factor uncertainties on top of any real anisotropy, with $\kappa_{22}:\kappa_{11}:\kappa_{33}\simeq1.00:1.43:5.30$ for a perfect match at $25$~K, see Fig.~\ref{Fig2}\textbf{b}. Here, the fact that $\kappa_{33}$ is about five times smaller than $\kappa_{11}$ and $\kappa_{22}$ reflects the quasi two-dimensional nature of the structure. Note that any contribution from mobile spin excitations, necessarily contained within the kagome planes and thus unable to carry heat along $c$, would preclude such scaling and would be substantially affected by the magnetic field. On the contrary, $\kappa_{22}$ and $\kappa_{33}$ are found to be field independent up to $15$~T while $\kappa_{11}$ displays a minute field dependence below $T_{\mathrm{S}1}$, see Fig.~\ref{Fig2}\textbf{a}. Most probably, the latter is related to the scattering of phonons by the paramagnetic spin fluctuations on a temperature range where short-range spin correlations (or paramagnons) start to develop\cite{Biesner_AQT_2022}. Increasing the magnetic field (here applied along $c$) gaps out some of these spin fluctuations, which in turn slightly enhances $\kappa_{11}$. 

It should also be pointed out that the three $\kappa_{\mathrm{ii}}$ are of remarkably modest magnitude, e.g., far below the low-temperature boundary scattering limit, and similar to the universal thermal conductivity of amorphous solids, see Fig.~\ref{Fig2}\textbf{b}. In particular, when considering the evolution of $\kappa_{\mathrm{ii}}$ with increasing temperature, a key feature is the slow and almost linear rise observed after a plateau-like regime ending at $T_{\mathrm{S}1}$. Such behaviour was explained at the theoretical level in terms of anharmonic interactions between phonons and fractons, which are short-scale vibrational excitations\cite{Alexander_PRB_1986}. Here, it seems logical to attribute the glass-like thermal conductivity to a strong scattering of phonons by the randomly distributed inter-plane hydrogen. While the latter freezes below $T_{\mathrm{S}1}$, $\kappa_{\mathrm{ii}}$ drops less rapidly with decreasing temperature: it remains more or less constant down to $T_{\mathrm{S}2}$, at which it experiences a minor improvement (most noticeable in $\kappa_{33}$) before vanishing.

\begin{figure}[t!]\centering
\includegraphics[width=\textwidth]{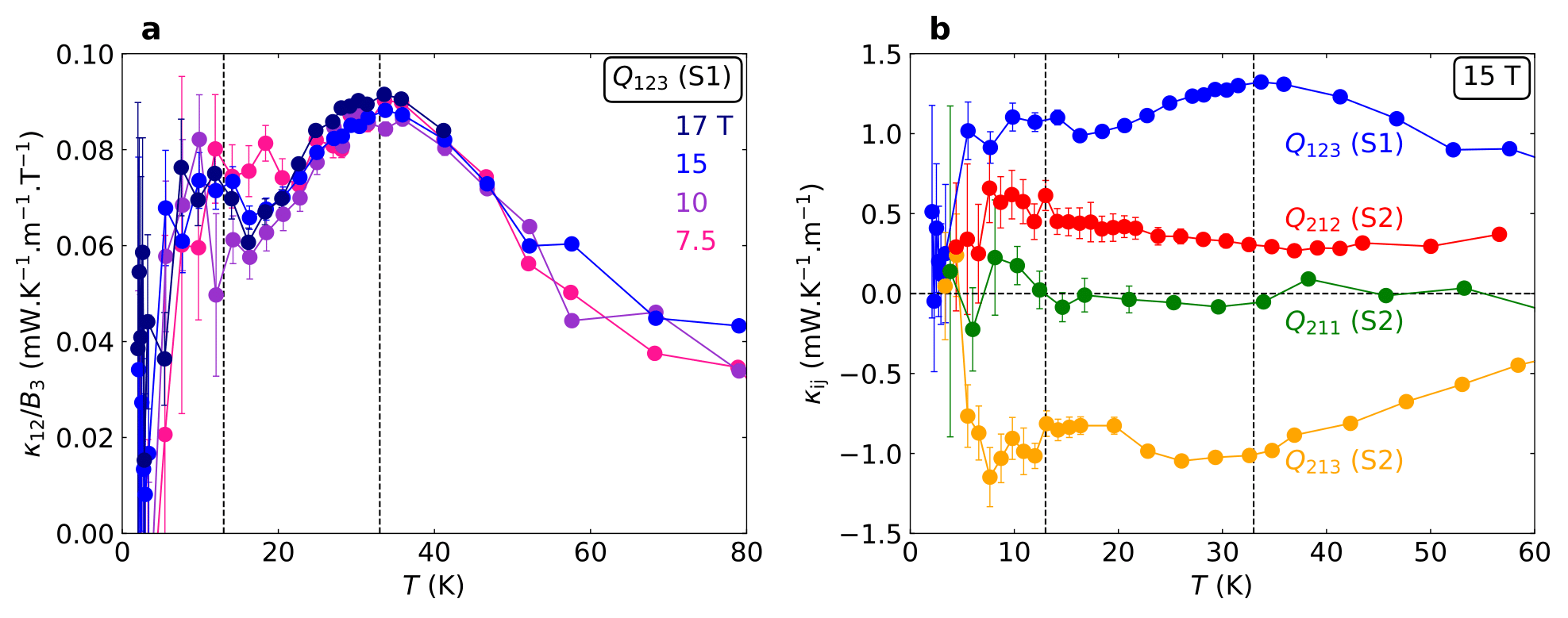}
\caption{\label{Fig3}\textbf{Conventional and planar parallel PHE in Y-kapellasite.} \textbf{a} Temperature dependence of the $Q_{123}$ Righi-Leduc component (measured on sample S1) plotted as $\kappa_{12}/B_{3}$ for several magnetic fields from $7.5$ to $17$~T (applied along $c$). \textbf{b} Temperature dependence of the four Righi-Leduc components $Q_{123}$ (conventional thermal Hall effect measured on sample S1), $Q_{213}$ (conventional thermal Hall effect measured on sample S2), $Q_{212}$ (planar parallel thermal Hall effect measured on sample S2) and $Q_{211}$ (planar orthogonal thermal Hall effect measured on sample S2), obtained in magnetic fields of $15$~T and plotted as $\kappa_{\mathrm{ij}}$. The two vertical black dashed lines in \textbf{a} and \textbf{b} indicate the structural transitions at $T_{\mathrm{S}1}\simeq33$~K and $T_{\mathrm{S}2}\simeq13$~K.}
\end{figure}

Now turning to the thermal Hall effect results presented in Fig.~\ref{Fig3}, we obtain small but finite $Q_{123}$, $Q_{213}$ and $Q_{212}$ components, while the $Q_{211}$ component is found to be virtually null over the whole temperature range. As for the $Q_{321}$ component, we were not able to obtain satisfactory data owing to the poor $\kappa_{33}$ which prevented us from generating any detectable thermal Hall gradient $\Delta T_{2}$ given our experimental sensitivity. In other words, a clear PHE is observed in two of the conventional configurations and in the planar parallel configuration. 

We first concentrate on the $Q_{123}$ component, which corresponds to the conventional thermal Hall conductivity $\kappa_{12}$, see Fig.~\ref{Fig3}\textbf{a}. It has a positive sign and a visible response to both structural transitions with a well-defined maximum position at $T_{\mathrm{S}1}$ and a fuzzier peak centred around $T_{\mathrm{S}2}$. Within reproducibility tolerance, it scales linearly with the magnetic field up to $17$~T, in stark contrast to expectations for the magnon Hall effect which declines with increasing magnetic field\cite{Onose_Science_2010}. This weak conventional PHE translates into a thermal Hall angle $\kappa_{12}/\kappa_{22}/B_{3}\simeq7\times10^{-5}$~T$^{-1}$ at $T_{\mathrm{S}1}$ in $15$~T. It is instructive to note that $\kappa_{12}$ and $\kappa_{22}$ have a different temperature dependence. While a shape similarity between $\kappa_{\mathrm{ij}}$ and $\kappa_{\mathrm{jj}}$ is often considered as an evidence that a single type of heat carriers (hence phonons) is involved\cite{Chen_arXiv_2023_a,Lefrancois_PRX_2022}, we demonstrate here that this criterion is a condition that may be sufficient but by no means necessary. For instance, when considering the evolution with decreasing temperature, $\kappa_{12}$ increases down to $T_{\mathrm{S}1}$ whereas $\kappa_{22}$ decreases. Then, down to $T_{\mathrm{S}2}$, $\kappa_{12}$ decreases more significantly than $\kappa_{22}$, which remains more or less constant. The latter decrease reveals that $\kappa_{12}$ is not enhanced by the positional freezing of the inter-plane hydrogen at low temperatures.

We now consider the $Q_{213}$ component, which corresponds to the conventional thermal Hall conductivity $\kappa_{21}$. According to the symmetry-adapted form of the Righi-Leduc tensor for the space groups $R\overline{3}$ and $P1$, $Q_{213}$ and $Q_{123}$ are Onsager-Casimir reciprocal, with $Q_{213}=-Q_{123}$, see Ext. Data Tables~\ref{Ext-Data-Table1}, \ref{Ext-Data-Table2}. As depicted in Fig.~\ref{Fig3}\textbf{b}, we confirm that $Q_{213}$ mirrors $Q_{123}$ with a negative sign and no significant magnitude difference. Note that this beautiful verification was carried out using two distinct samples, which underlines the reliability and reproducibility of our measurements.

Finally, we discuss the two most exciting components $Q_{212}$ and $Q_{211}$, which correspond to the planar parallel and planar orthogonal thermal Hall conductivities $\kappa_{21}$ when the magnetic field is applied either parallel or orthogonal to the heat flux within the $(ab)$ plane. We immediately rule out a spurious occurrence of $Q_{212}$ arising from a contamination by $Q_{213}$. On the one hand, $Q_{212}$ is found to have the opposite sign, positive, and displays a different temperature dependence which may in turn suggest a different origin, see Fig.~\ref{Fig3}\textbf{b}. In particular, we notice that $Q_{212}$ increases with decreasing temperature from $T_{\mathrm{S}1}$ to $T_{\mathrm{S}2}$ whereas $|Q_{213}|$ decreases like $Q_{123}$. On the other hand, we took great care to check the alignment of the magnetic field and estimate that it deviates from the $(ab)$ plane towards $c$ by $\pm5$~\textdegree~at most. This corresponds to a maximal contamination of about $9$~\% of $|Q_{213}|$, well below the observed $Q_{212}$, with for instance $0.09\times0.81\simeq0.07$~mW.K$^{-1}$.m$^{-1}$ versus $0.61$~mW.K$^{-1}$.m$^{-1}$ (more than eight times larger) at $T_{\mathrm{S}2}$. 

Detecting a finite $Q_{212}$ at temperatures above $T_{\mathrm{S}1}$ comes as a real surprise because it is forbidden in the symmetry-adapted form of the Righi-Leduc tensor for the space group $R\overline{3}$, see Ext. Data Tables~\ref{Ext-Data-Table1}. Its existence necessarily implies the occurrence of special symmetry breakings. We therefore listed all the space groups resulting from lowering the symmetries of $R\overline{3}$ down to $P1$, see Ext. Data Fig.~\ref{Ext-Data-Fig1}. Among all possible subgroups, only $P\overline{1}$ and $P1$ have symmetries compatible with a finite $Q_{212}$ (note that a finite $Q_{211}$ is also allowed), see Ext. Data Tables~\ref{Ext-Data-Table1}, \ref{Ext-Data-Table2}. Therefore, in contrast to the cases of $Q_{123}$ and $Q_{213}$, the inter-plane hydrogen may play a crucial role in the establishment of $Q_{212}$. We propose that a local symmetry breaking from $R\overline{3}$ to $P\overline{1}$ or $P1$ resulting from the random distribution of this atom is sufficient for the planar parallel PHE to emerge above $T_{\mathrm{S}1}$. It is then only slightly enhanced when some if not all of the broken symmetries become global below $T_{\mathrm{S}1}$ and peaks around $T_{\mathrm{S}2}$, see Fig.~\ref{Fig3}\textbf{b}. 

For now, it is also not clear why $Q_{211}$ remains vanishingly small but this observation, in line with other planar thermal Hall effect studies\cite{Yokoi_Science_2021,Czajka_NM_2022,Chen_arXiv_2023_a}, may prove useful in subsequent theoretical developments to clarify the precise mechanisms responsible for the planar PHE. For comparison, $Q_{\mathrm{iji}}$ and $Q_{\mathrm{ijj}}$ are correspondingly reported to be finite and null in both $\alpha$-RuCl$_{3}$ (in agreement with the symmetries of the space group $C2/m$) and Na$_{2}$Co$_{2}$TeO$_{6}$ (although a finite $Q_{\mathrm{iji}}$ is there forbidden by the symmetries of the space group $P6_{3}22$).

In summary, our thermal transport measurements on Y-kapellasite yield a paradigm shift in the study of the thermal Hall effect in insulators. First, we report the first unambiguous detection of a planar parallel PHE. Our findings thus prompt a second look at the interpretations put forward in previous studies on quantum spin liquid candidates and open up a wider range of scenarios in which the phonon contribution cannot be neglected. Second, we now have a specific and concrete case for the general idea of local symmetry breaking at the root of the PHE in a wide range of insulators.

\clearpage

{\small

\begin{methods}

\paragraph{Structure and magnetic model of Y-kapellasite.}{Y-kapellasite is a derivative of the emblematic quantum spin liquid candidate herbertsmithite ZnCu$_{3}$(OH)$_{6}$Cl$_{2}$, discovered as an interesting by-product of unsuccessful doping attempts when substituting divalent zinc for trivalent yttrium\cite{Puphal_JMCC_2017,Barthelemy_PRM_2019}. In this material, copper spins $S=1/2$ decorate a slightly distorted kagome lattice with yttrium located close to the centre of hexagons, see Fig.~\ref{Fig1}\textbf{b}, hence the kapellasite denomination (Zn-kapellasite is a polymorph of herbertsmithite, the difference between the two structures being the position of zinc, either within or in between the kagome planes). Contrary to zinc, yttrium has a significantly larger ionic radius than copper, which precludes any intersite mixing and renders the system immune to the troublesome magnetic defects typical of herbertsmithite, Zn-kapellasite or Zn-barlowite. Hydroxyl groups and chlorine constitute thick diamagnetic layers separating the kagome planes and recent \emph{ab initio} density functional theory combined with inelastic neutron scattering results confirmed the quasi two-dimensional nature of the magnetic lattice, demonstrating that three intra-plane antiferromagnetic Heisenberg couplings between nearest neighbours dominate over all other possible intra- or inter-plane couplings between further neighbours\cite{Hering_NPJCM_2022,Chatterjee_PRB_2023}. These main three couplings $J\simeq J_{\hexagon}\simeq140$~K and $J^{'}\simeq63$~K result in an original anisotropic variant of the standard nearest neighbour Heisenberg model (recovered for $J=J_{\hexagon}=J^{'}$), breaking translational symmetry of the kagome lattice but retaining six-fold rotational symmetry around hexagons, see Fig.~\ref{Fig1}\textbf{b}. Owing to the considerable frustration produced by the lattice geometry and the competition between the latter three antiferromagnetic terms, the material remains paramagnetic down to $T_{\mathrm{N}}\simeq2$~K, below which a coplanar long-range order with propagation vector $Q=(1/3,1/3)$ sets in, as predicted theoretically for the ground state\cite{Hering_NPJCM_2022}. This magnetic transition, resulting in a remarkably weak ordered moment of about $1/30$~$\mu_{\mathrm{B}}$, initially remained elusive when focusing on polycrystalline samples\cite{Barthelemy_PRM_2019} until it was later demonstrated in the case of large phase-pure single crystals prepared via optimal synthesis\cite{Chatterjee_PRB_2023}. For that matter, every improvement in the synthesis procedure triggered a thorough reappraisal of the exact stoichiometry and crystallographic structure which were eventually settled through inductively coupled plasma mass spectroscopy, gas extraction and detailed neutron diffraction measurements from $40$~K down to $65$~mK\cite{Chatterjee_PRB_2023}. Y-kapellasite crystallises in a trigonal rhombohedral structure defined by the space group $R\overline{3}$ ($148$), in which the inter-plane hydrogen randomly occupies six equivalent positions, thus locally breaking the global crystal symmetry, see Fig.~\ref{Fig1}\textbf{b}. Upon cooling, two structural transitions occuring at $T_{\mathrm{S}1}\simeq33$~K and $T_{\mathrm{S}2}\simeq13$~K were detected through specific heat, thermal expansion and $^{35}$Cl NMR measurements on single crystals while they remained unnoticed in prior studies of polycrystalline samples\cite{Barthelemy_PRM_2019}. Strikingly, these transitions only reflect in the neutron diffraction data through a clear intensity increase for some Bragg peaks. Preserving the same space group and lattice parameters down to the lowest temperatures does not affect the refinement quality. The transition at $T_{\mathrm{S}1}$ is attributed to the positional freezing of the inter-plane hydrogen, thereby potentially leading to a global crystal symmetry breaking from $R\overline{3}$ to $P1$ ($1$) with three different twin domains. This is compatible with the complex $^{35}$Cl quadrupolar line splitting reported below $T_{\mathrm{S}1}$. Further terahertz magnetometry measurements revealed that these structural transitions are accompanied by the building up of short-range spin correlations (or paramagnons) although long-range magnons only emerge below $T_{\mathrm{N}}$ as highlighted with inelastic neutron scattering\cite{Biesner_AQT_2022,Chatterjee_PRB_2023}.}

\paragraph{Optimal synthesis of Y-kapellasite.}{The crystal growth of Y-kapellasite was originally reported in Reference\cite{Puphal_JMCC_2017}. Subsequently, as described in Reference\cite{Biesner_AQT_2022}, the synthesis was improved to obtain inclusion-free, large, bulk single crystals by means of a horizontal external gradient method in thick-walled quartz ampoules with a wall thickness of $2.5$-$3$~mm. Growth is achieved by slowly dissolving CuO in a YCl$_{3}$-H$_{2}$O solution and transporting it to the cold end. This is realised in a three-zone furnace with a gradient of $25$~\textdegree C and a temperature of $240$~\textdegree C at the hot end, over a length of $20$~cm. The gradient was optimised because too low temperatures yielded a phase mixture of Y-kapellasite and clinoatacamite. The phase-pure, optically transparent single crystals have an average size of $3\times3\times1$~mm$^{3}$ up to $3\times3\times3$~mm$^{3}$ when grown over several weeks. Their orientation is facilitated by their hexagonal plaquette shape, with $c$ perpendicular to the hexagonal faces and $a$ (respectively $b^{\star}$) perpendicular (respectively parallel) to the hexagonal edge. The samples S$1$, S$2$ and S$3$ examined here are from the same batch as the single crystals investigated in References\cite{Chatterjee_PRB_2023,Biesner_AQT_2022}. S$1$ and S$2$ were selected among the thinnest and measured as grown, while S$3$ was cut along $c$ in one of the thickest.}

\paragraph{Thermal transport measurements.}{Thermal transport measurements were performed using a standard steady-state method. A constant heat flux $q_{i}$ is injected at one end of the sample along the direction $i$ while the other end is thermally sunk to a heat bath at temperature $T_{0}$ (either a copper or lithium fluoride block), see Fig.~\ref{Fig1}\textbf{a}. The heat flux is generated by applying an electric current through a strain gauge whose resistance (of about $5$~k$\Omega$) marginally depends on the temperature and magnetic field. Assuming a one-dimensional heat flow and an isotropic medium, the longitudinal thermal gradient $\Delta T_{\mathrm{i}}$ is measured between two contacts separated by a distance $l$ along the direction $i$, see Fig.~\ref{Fig1}\textbf{a}. This gradient is assessed using either two Cernox sensors (calibrated \emph{in situ} against a reference Cernox) or two type-E thermocouples. The longitudinal thermal conductivity $\kappa_{\mathrm{ii}}$ is given by $\kappa_{\mathrm{ii}}=q_{i}/(\alpha\Delta T_{\mathrm{i}})$, where $\alpha$ is a geometric factor determined by the cross section $wt$ ($w$: width, $t$: thickness) divided by $l$. In presence of an applied magnetic field $B_\mathrm{x}$, with $x\in\{i,j,k\}$, the orthogonal thermal gradient $\Delta T_{\mathrm{j}}$ is measured between two contacts separated by a distance $w$ along the direction $j$, see Fig.~\ref{Fig1}\textbf{a}. This gradient is assessed using a differential type-E thermocouple and antisymmetrised between the two field polarities to remove any contamination by the longitudinal gradient: $\Delta T_{\mathrm{j}}(B_\mathrm{x})=[\Delta T_{\mathrm{j}}(+B_\mathrm{x})-\Delta T_{\mathrm{j}}(-B_\mathrm{x})]/2$. The thermal Hall conductivity $\kappa_{\mathrm{ij}}$ is given by $\kappa_{\rm ij}=l\kappa_{\mathrm{jj}}\Delta T_{\mathrm{j}}/(w\Delta T_{\mathrm{i}})$, which implies a two-step computation because $\kappa_{\mathrm{jj}}$ and $\Delta T_{\mathrm{j}}$ are not measured simultaneously. Error bars in the figures represent one standard deviation. In our mountings, all connections between the samples and the heat baths, temperature sensors and strain gauges consisted of gold and silver wires with a $17$ to $100$~$\mu$m diameter attached using silver paste. The contacts had geometries ($l\times w\times t$) $1859(60)\times2239(78)\times334(10)$~$\mu$m$^{3}$ on S$1$, $558(108)\times1040(60)\times85(4)$~$\mu$m$^{3}$ on S$2$ and $418(70)\times600(99)\times137(38)$~$\mu$m$^{3}$ on S$3$. Note that a quantitative determination of $\kappa_{\mathrm{ij}}$ requires measuring both $\kappa_{\mathrm{jj}}$ and $\Delta T_{\mathrm{j}}$ in the same sample. Here, the longitudinal thermal conductivity $\kappa_{11}$ and the thermal Hall gradient $\Delta T_{2}$ were measured in sample S$1$ while the longitudinal thermal conductivity $\kappa_{22}$ and the thermal Hall gradient $\Delta T_{1}$ were measured in sample S$2$. Owing to the modest anisotropy between $\kappa_{11}$ and $\kappa_{22}$ (if any beyond potential variations in crystalline quality and geometric factor uncertainties), see Fig.~\ref{Fig2}, we assumed that $\kappa_{11}\simeq\kappa_{22}$ to compute $\kappa_{21}$ so as to compare $\kappa_{12}$ and $\kappa_{21}$ in a meaningful way.}

\end{methods}

}

\clearpage

\begin{addendum}
\item We are grateful to S. Fortier for extensive technical support and acknowledge valuable discussions with N. Gauthier and J.A. Quilliam. C.K. acknowledges funding from the Deutsche Forschungsgemeinschaft (DFG) through TRR $288$-$422213477$ (project A$03$). L.T. acknowledges support from the Canadian Institute for Advanced Research (CIFAR) as a Fellow and funding from the Institut Quantique, the Natural Sciences and Engineering Research Council of Canada (NSERC, PIN $123817$), the Fonds de Recherche du Québec - Nature et Technologies (FRQNT), the Canada Foundation for Innovation (CFI), and a Canada research chair. This research was undertaken thanks in part to funding from the Canada First Research Excellence Fund.
\item[Author contributions] Q.B., P.P. and L.T. conceived and led the project. P.P., K.M.Z. and C.K. grew the single crystals. Q.B., É.L., L.C., and A.V. carried out the thermal transport measurements and analysis. Q.B. wrote the manuscript with feedback from all the authors.    
\item[Competing interests] The authors declare no competing interests.
\item[Correspondence] Correspondence and requests for materials should be addressed to Q.B., P.P. or L.T.
\end{addendum}

\clearpage

\renewcommand{\figurename}{Fig.}

\clearpage

\clearpage

\clearpage

\begin{sidewaystable}\centering
\begin{tabular}{|c|c|c|c|c|c|}
\hline
Configuration & Sample & Heat flux $q_{\mathrm{i}}$ & Thermal Hall gradient $\Delta T_{\mathrm{j}}$ & Applied magnetic field $B_{\mathrm{x}}$ & Righi-Leduc tensor component $Q_{\mathrm{ijx}}$ \\
\hline
$1$ & S$1$ & $\parallel a(1)$ & $\parallel b^{\star}(2)$ & $\parallel c(3)$ & $Q_{123}$ \\
\hline
$2$ & S$2$ & $\parallel b^{\star}(2)$ & $\parallel a(1)$ & $\parallel c(3)$ & $Q_{213}=-Q_{123}$ \\
\hline
$3$ & S$2$ & $\parallel b^{\star}(2)$ & $\parallel a(1)$ & $\parallel b^{\star}(2)$ & $Q_{212}$ \\
\hline
$4$ & S$2$ & $\parallel b^{\star}(2)$ & $\parallel a(1)$ & $\parallel a(1)$ & $Q_{211}$ \\
\hline
$5$ & S$3$ & $\parallel c(3)$ & $\parallel b^{\star}(2)$ & $\parallel a(1)$ & $Q_{321}$ \\
\hline
\end{tabular}
\caption{\label{Table1}\textbf{Details of the five thermal transport configurations that were adopted.} For each configuration, the sample, the directions of the heat flux, the thermal Hall gradient and the applied magnetic field are specified along with the corresponding Righi-Leduc tensor component. Conventional thermal Hall effects were investigated in samples S$1$, S$2$ and S$3$ using configurations $1$, $2$ and $5$ respectively. Planar parallel and planar orthogonal thermal Hall effects were investigated in sample S$2$ using configurations $3$ and $4$ respectively.}
\end{sidewaystable}

\clearpage

\renewcommand{\figurename}{Extended Data Fig.}
\setcounter{figure}{0}

\begin{figure}[t!]\centering
\includegraphics[width=0.8\textwidth]{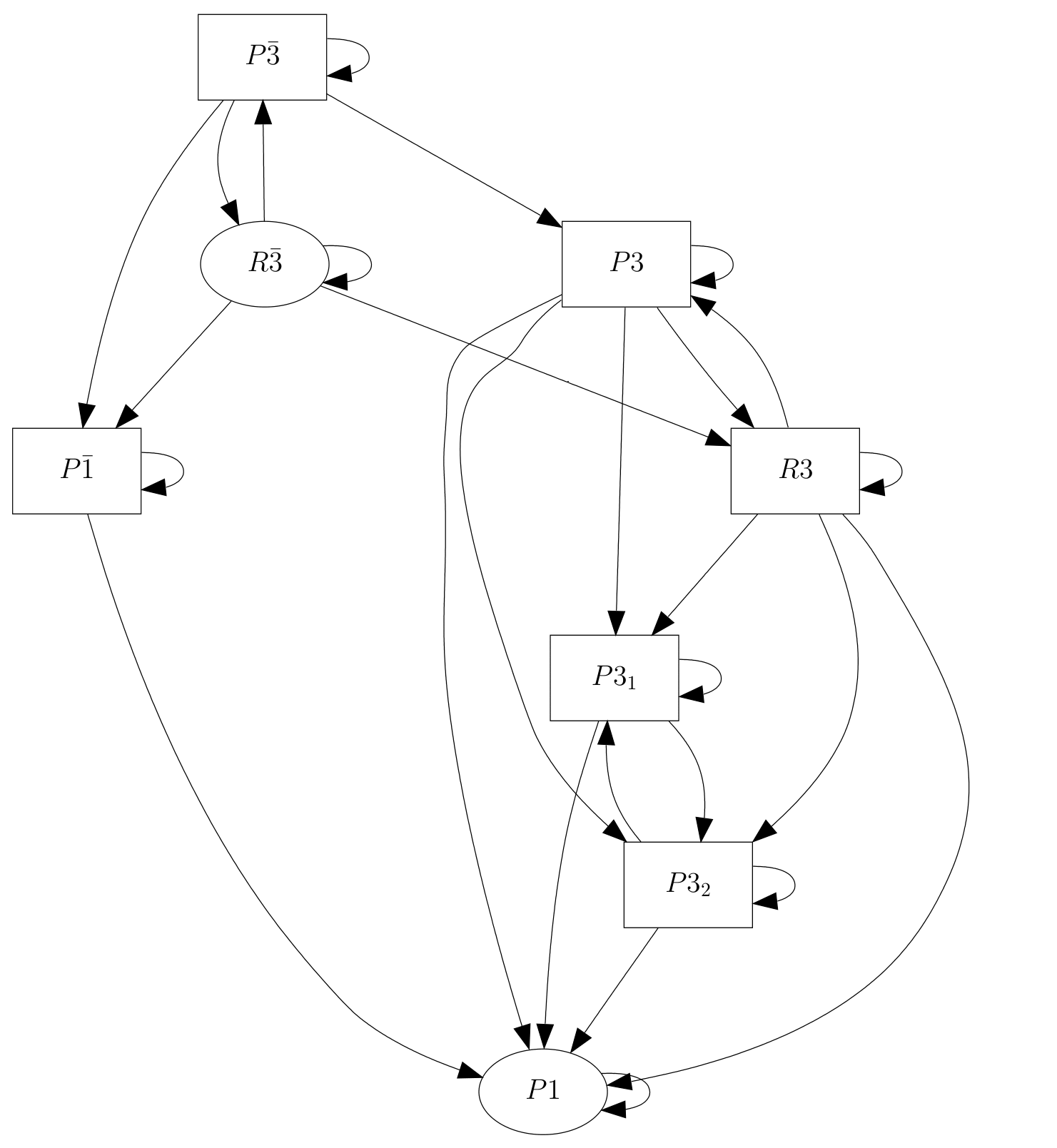}
\caption{\label{Ext-Data-Fig1}\textbf{Graph of maximal subgroups from $\mathbf{R\overline{3}}$ ($\mathbf{148}$) to $\mathbf{P1}$ ($\mathbf{1}$).} The space groups $P\overline{3}$ ($147$), $R3$ ($146$), $P3_{2}$ ($145$), $P3_{1}$ ($144$), $P3$ ($143$), $P\overline{1}$ ($2$) and $P1$ are obtained by gradually breaking down the symmetries of $R\overline{3}$. This graph was produced using the SUBGROUPGRAPH module of the Bilbao Crystallographic Server\cite{Ivantchev_JAC_2000}.}
\end{figure}

\clearpage

\renewcommand{\tablename}{Extended Data Table}
\setcounter{table}{0}

\begin{table}\centering
\includegraphics[width=0.5\textwidth]{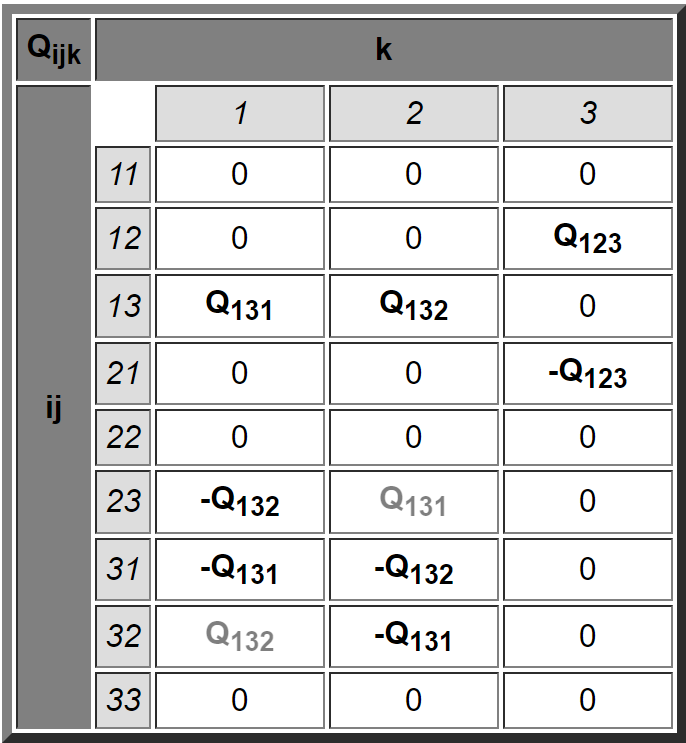}
\caption{\label{Ext-Data-Table1}\textbf{Symmetry-adapted form of the Righi-Leduc tensor $\mathbf{Q_{ijk}}$ for the space groups $\mathbf{R\overline{3}}$ ($\mathbf{148}$), $\mathbf{P\overline{3}}$ ($\mathbf{147}$), $\mathbf{R3}$ ($\mathbf{146}$), $\mathbf{P3_{2}}$ ($\mathbf{145}$), $\mathbf{P3_{1}}$ ($\mathbf{144}$) and $\mathbf{P3}$ ($\mathbf{143}$).} Finite $Q_{123}$ and $Q_{213}=-Q_{123}$ components, corresponding to conventional thermal Hall effects, are allowed. Finite $Q_{211}$ and $Q_{212}$ components, corresponding to planar orthogonal and planar parallel thermal Hall effects, are forbidden. This table was produced using the TENSOR module of the Bilbao Crystallographic Server\cite{Gallego_ACA_2019}.}
\end{table}

\clearpage

\begin{table}\centering
\includegraphics[width=0.5\textwidth]{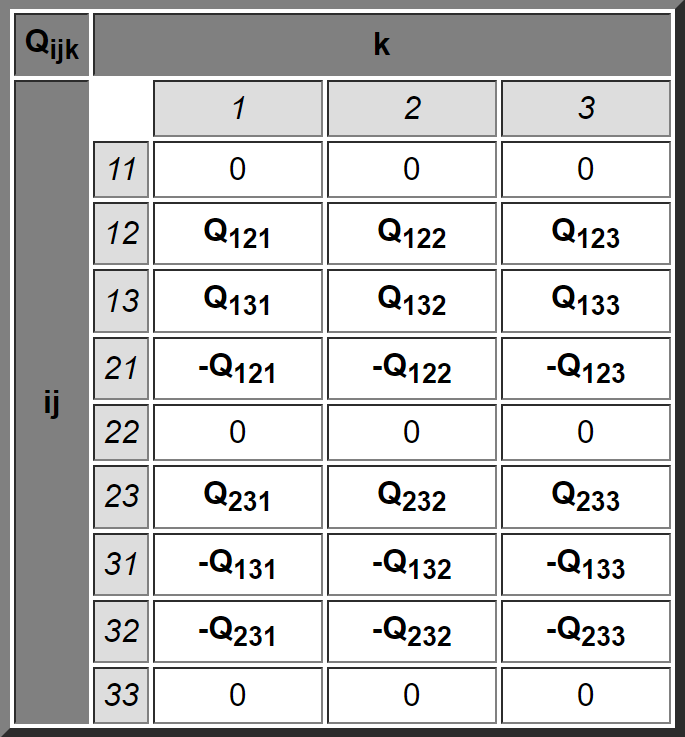}
\caption{\label{Ext-Data-Table2}\textbf{Symmetry-adapted form of the Righi-Leduc tensor $\mathbf{Q_{ijk}}$ for the space groups $\mathbf{P\overline{1}}$ ($\mathbf{2}$) and $\mathbf{P1}$ ($\mathbf{1}$).} Finite $Q_{123}$ and $Q_{213}=-Q_{123}$ components, corresponding to conventional thermal Hall effects, are allowed. Finite $Q_{211}=-Q_{121}$ and $Q_{212}=-Q_{122}$ components, corresponding to planar orthogonal and planar parallel thermal Hall effects, are also allowed. This table was produced using the TENSOR module of the Bilbao Crystallographic Server\cite{Gallego_ACA_2019}.}
\end{table}

\clearpage


\clearpage

\begin{thebibliography}{1}

\bibitem{Yokoi_Science_2021} T. Yokoi \emph{et al.}, \emph{Science} 373, 6554, 568-572 (2021) 
\bibitem{Czajka_NM_2022} P. Czajka \emph{et al.}, \emph{Nat. Mater.} 22, 36-41 (2023)
\bibitem{Takeda_PRR_2022} H. Takeda \emph{et al.}, \emph{Phys. Rev. Res.} 4, L042035 (2022)
\bibitem{Chen_arXiv_2023_a} L. Chen \emph{et al.}, \emph{arXiv}:2309.17231 (2023)
\bibitem{Chen_arXiv_2023_b} L. Chen \emph{et al.}, \emph{arXiv}:2310.07696 (2023)
\bibitem{Strohm_PRL_2005} C. Strohm \emph{et al.}, \emph{Phys. Rev. Lett.} 95, 155901 (2005)
\bibitem{Kitaev_AP_2006} A. Kitaev, \emph{Ann. Phys.} 321, 1, 2-111 (2006)
\bibitem{Katsura_PRL_2010} H. Katsura \emph{et al.}, \emph{Phys. Rev. Lett.} 104, 066403 (2010)
\bibitem{Onose_Science_2010} Y. Onose \emph{et al.}, \emph{Science} 329, 5989, 297-299 (2010)
\bibitem{Hirschberger_PRL_2015} M. Hirschberger \emph{et al.}, \emph{Phys. Rev. Lett.} 115, 106603 (2015)
\bibitem{Gao_SciPost_2020} Y. H. Gao and G. Chen, \emph{SciPost Phys. Core} 2, 004 (2020)
\bibitem{Teng_PRR_2020} Y. Teng \emph{et al.}, \emph{Phys. Rev. Res.} 2, 033283 (2020)
\bibitem{Chern_PRL_2021} L. E. Chern \emph{et al.}, \emph{Phys. Rev. Lett.} 126, 147201 (2021)
\bibitem{Lefrancois_PRX_2022} É. Lefrançois \emph{et al.}, \emph{Phys. Rev. X} 12, 021025 (2022) 
\bibitem{Gillig_arXiv_2023} M. Gillig \emph{et al.}, \emph{arXiv}:2303.03067 (2023)
\bibitem{Chatterjee_PRB_2023} D. Chatterjee \emph{et al.}, \emph{Phys. Rev. B} 107, 125156 (2023)
\bibitem{Biesner_AQT_2022} T. Biesner \emph{et al.}, \emph{Adv. Quantum Technol.} 5, 2200023 (2022)
\bibitem{Alexander_PRB_1986} S. Alexander \emph{et al.}, \emph{Phys. Rev. B} 34, 2726 (1986)
\bibitem{Puphal_JMCC_2017} P. Puphal \emph{et al.}, \emph{J. Mater. Chem. C} 5, 2629-2635 (2017)
\bibitem{Barthelemy_PRM_2019} Q. Barthélemy \emph{et al.}, \emph{Phys. Rev. Mater.} 3, 074401 (2019)
\bibitem{Hering_NPJCM_2022} M. Hering \emph{et al.}, \emph{npj Comput. Mater.} 8, 10 (2022)
\bibitem{Ivantchev_JAC_2000} S. Ivantchev \emph{et al.}, \emph{J. Appl. Cryst.} 33, 1190-1191 (2000)
\bibitem{Gallego_ACA_2019} S.V. Gallego \emph{et al.}, \emph{Acta Cryst. A} 75, 438-447 (2019)
\end{thebibliography}
\end{document}